# Impact of Resonant Magnetic Perturbations on the

# L-H Transition on MAST

R. Scannell[1], A. Kirk[1], M. Carr[1], J. Hawke[2], S. S. Henderson[3], T. O'Gorman[1], A. Patel[1], A. Shaw[1], A. Thornton[1]and the MAST Team

[1]CCFE, Culham Science Centre, Abingdon, Oxon OX14 3DB, UK

[2]FOM Institute DIFFER, Dutch Institute for Fundamental Energy Research, Association EURATOM-FOM, 3430 BE Niewegein, Netherlands

[3]Department of Physics, SUPA, University of Strathclyde, Glasgow, G4 0NG

E-mail: rory.scannell@ccfe.ac.uk

**Abstract** The impact of resonant magnetic perturbations (RMPs) on the power required to access H-mode is examined experimentally on MAST. Applying RMP in n=2,3,4 and 6 configurations causes significant delays to the timing of the L-H transition at low applied fields and prevents the transition at high fields. The experiment was primarily performed at RMP fields sufficient to cause moderate increases in ELM frequency, $f_{mitigated}/f_{natural} \sim 3$. To obtain H-mode with RMPs at this field, an increase of injected beam power is required of at least 50% for n=3 and n=4 RMP and 100% for n=6 RMP. In terms of power threshold, this corresponds to increases of at least 20% for n=3 and n=4 RMPs and 60% for n=6 RMPs. This 'RMP affected' power threshold is found to increase with RMP magnitude above a certain minimum perturbed field, below which there is no impact on the power threshold. Extrapolations from these results indicate large increases in the L-H power threshold will be required for discharges requiring large mitigated ELM frequency.

## 1. Introduction

The power loading to the divertor during type I ELMs is a concern for large fusion machines [1]. One proposed method to reduce this power loading is the application of resonant magnetic perturbations (RMPs). These RMPs have been used to suppress type I ELMs as demonstrated on DIII-D[2], ASDEX Upgrade[3] and KSTAR[4], however, not all devices with RMPs have achieved suppression. A second possibility is mitigation, which reduces the power loading due to the ELMs, typically by increasing ELM frequency and hence reducing the energy loss per ELM event. Mitigation is observed on a number of tokamaks, notably MAST[5] and JET[6, 7]. A further possible reason for RMPs in future devices is that ELMs may be required at a certain minimum frequency to prevent Tungsten accumulation [8] by increasing particle transport at the edge.

One observed side effect of the application of RMPs is that it makes H-mode access more difficult increasing the L-H Power threshold ($P_{L-H}$) and hence the requirement for external heating power in order to access H-mode. This increased power requirement is small in absolute terms on current devices. The power threshold scales approximately as $P_{L-H}=0.0488n_{e20}^{0.717}S^{0.94}B_T^{0.8}$ [9], where $n_{e20}$ is the line average density in $10^{20}m^{-3}$, $B_T$ is the magnetic field and S the plasma surface area. Extrapolating using this scaling, the power threshold for ITER is predicted to be ~52MW[10]. Increases in the ITER power threshold in the presence of RMP similar to those observed on current devices could potentially be quite large. One positive development is that recent results with the full metal wall on ASDEX Upgrade[11] and JET[12] have indicated that the power threshold requirement may be reduced by 25-30% with respect to operation with the Carbon wall. However, the L-H power threshold on ITER in the presence of RMP remains a concern particularly due to the uncertainty in the parameters which influence the magnitude of this increased $P_{L-H}$.

On DIII-D [13] it has been observed that in Deuterium discharges the $P_{L-H}$ increases once the magnitude of the RMP goes above a critical threshold. Increases in power threshold of up to 100% have been observed at large field perturbation $\delta B/B_T$. Experiments on the impact of RMPs on $P_{L-H}$ on ASDEX upgrade in n=2 [14] and [9] have been performed showing a 20% increase in $P_{L-H}$ for $0.45n_{GW}<n_e<0.65n_{GW}$. In these discharges, the transition during RMP is observed to occur at a higher, less negative, radial electric field minimum. On NSTX [15] an increase in $P_{L-H}$ of at least 50% is observed on application of RMP. The NSTX results again points to a possible dependence of the increase in $P_{L-H}$ on changes in the radial electric field magnitude.

On MAST [16], application of n=3 RMP to a 900kA double null discharge results in an increase of required beam power from 1.8MW to 3.3MW to achieve a H-mode transition at the same time as a no applied RMP discharge. In discharges prevented from accessing H-mode by application of RMP, a more positive Lorentz (v×B) component of the radial electric field, as measured from Doppler spectrometry from Helium, is observed. The study that will be presented in this paper differs from [16] in that the resulting plasmas exhibit useful ELM mitigation as a result of the applied RMPs. In addition, the discharges examined in this paper are lower single null discharges as opposed to connected double null, and hence more similar to the ITER shape.

For a more general overview of the results of application of RMPs to MAST discharges, the reader is referred to [17] and the references therein. An investigation into the H-mode power threshold scaling with plasma parameters for a number of machines is described in [18]. The power threshold on MAST has been previously investigated in [19], specifically the variation of the power threshold with separatrix configuration [20] and divertor leg length [21].

The format of this paper is as follows. Section 2 examines the impact of RMP on a 600kA discharge with constant fuelling. Section 3 discusses the natural L-H power threshold for 400kA discharges which are the focus for the remainder of this paper. Section 4 looks at the results of an RMP field scan on L-H transition and examines the plasma profiles before the transition. Finally, in section 5, the impact of different toroidal mode number 'n' RMP configurations on the L-H power threshold are compared and quantified.

## 2. Impact of RMP on L-H transition at constant fuelling (600kA n=6)

In this section the impact of an applied n=6 RMP field on the L-H transition of a lower single null discharge with a plasma current ($I_p$) of 600kA is examined. The discharges had constant gas fuelling rate, a toroidal field on axis of 0.55T and a safety factor ($q_{95}$) of 4.0. To determine the power threshold for similar discharges without RMP a dedicated scan in NBI power was performed. This scan showed that 0.6MW or greater of applied neutral beam power causes a transition to H-mode. The duration of the plasma current flat top decreases with decreasing neutral beam power and in shots with 0.4MW or less of injected power the plasma current begins to fall off before the L-H transition time. Hence the L-H power threshold is at most 0.6MW, but could be lower. This scan was performed at a single line integral density and the variation of the L-H power threshold with density for these lower single null discharges has not been examined. The discharge to which n=6 RMP has been applied had 1.5MW of injected neutral beam power and so is well above the L-H power threshold.

The midplane $D\alpha$ emission for the no applied RMP and discharges with 1.0kA and 1.4kA in the RMP coils are shown in figure 1(a)-(c). The small spikes in the $D\alpha$ measurements during the L-mode are indicative of sawteeth which are a feature of these discharges. The L-H transitions in all three discharges shown are induced by sawteeth and hence co-incident with the sawtooth event. With no applied RMP the L-H transition occurs at 330ms. Application of RMP with 1.0kA of current in the coils delays the transition to 375ms (the subsequent spike in density at 400ms is due to injection of a pellet). Application of RMP with 1.4kA of current in the coils completely suppresses the transition during the plasma current flat top. A subsequent L-H transition in this discharge observed at 420ms, however, this transition is induced by the plasma current ramp down and hence not considered to be a 'natural' transition. The timing of the $I_p$ ramp down causes the gas to turn off, hence is at the same time as the reduction in gas flow rate as shown in figure 1(d).

In these discharges the gas flow rate was kept constant from 250ms. Upon applying RMP there is a particle pump out. However, the L-H transition occurs at similar measured line integral density for all three discharges. The result in figure 1 for constant gas flow rate illustrates that the suppression and delay of the L-H transition is due to application of RMP field and not due to change of edge conditions as a result of changing fuelling. In all other discharges discussed in this paper the line integral density is fixed by feedback control on the gas flow in order that the impact on the L-H transition may be observed at constant density. In these discharges with feedback control the RMP field causes a density loss that scales with RMP intensity and hence the gas flow rate increases with increasing RMP field.

The final trace shown in figure 1(f) is the neutron rate as measured by a $U^{235}$ fission chamber [22] which is similar for all 3 discharges. The neutron rate in these discharges are predominantly due to interaction of injected neutral beam Deuterium with fast ions in the plasma, hence a variation in fast ion loss rate would be expected to change this neutron rate. Since the neutron rates are similar for the discharges shown here, in this case there is no evidence that the RMPs impact on the fast ion confinement.

### 3. Power Threshold for 400kA Discharges

The impact of RMPs on ELM frequency in MAST plasmas has been widely studied in lower single null discharges with plasma currents of both 600kA and 400kA. Lower single null discharges are chosen as this is the operating configuration for ITER. In the following sections RMP are applied to shots in 400kA discharges. These 400kA shots were chosen over the 600kA discharge for this study because: 1) the 400kA discharges have fewer sawteeth, hence the triggering of the L-H transition by sawteeth events is less likely 2) the natural ELM frequency and increase in ELM frequency due to application of RMP is easier to observe without the sawtooth induced ELMs triggered in 600kA discharges and 3) a longer plasma current flat-top duration can be maintained in 400kA discharges hence allowing longer time for a delayed L-H transition to occur.

The results of a scan in neutral beam power to determine the power threshold in these 400kA discharges without RMPs is shown in figure 2. The shots are similar with beam power applied from 130ms (except for beam breakdown in the 1.5MW discharge which marginally delaying the start time) and the discharges go into density feedback from 310ms such that the L-H transition in all discharges occurs at the same density. The reduction of beam power from 1.5MW to 1.2MW and then to 0.9MW does not affect the time of the L-H transition. At 0.6MW of injected power a delayed transition occurs, followed by a back transition and a subsequent L-H transition. This scan indicates the L-H power threshold for no applied RMP in 400kA discharges corresponds to an injected beam power of 0.6MW or less. No lower beam power was available to obtain a discharge under the L-H power threshold, however, due to the weakness of the resulting H-mode obtained it is likely that the threshold is close to 0.6MW of injected power.

An estimated loss power $P_{Loss} = P_{NBI} + P_{OHMIC} - dW/dt - X$ is also shown in figure 2, where $P_{NBI}$ is the injected beam power, $P_{OHMIC}$ the product of loop voltage and plasma current, $dW/dt$ represents the change in plasma energy and X the change in stored magnetic energy. There is a variation in $P_{Loss}$ due to gas fuelling and during the H-mode due to changes in plasma stored energy between ELMs and sawteeth. The values of $dW/dt$ and X are obtained from the equilibrium reconstruction code EFIT [23]. However, this $P_{Loss}$ estimate has two shortcomings, it uses injected beam power as opposed to absorbed beam power and does not take into account loss of power due to radiation. Absorbed beam powers of ~60% are estimated for the L-mode phases of these discharges from power accounting. The shinethrough is relatively high as the plasma is ~20cm below the vertical midplane of the vessel where the beams are injected. The power threshold, $P_{th}$, is the minimum value of the particle power flowing through the separatrix ($P_{Sep}$) required to cause an L-H transition. The value $P_{Sep}$ could be obtained from $P_{Loss}$ by making the adjustments $P_{Sep}=P_{Loss}-P_{Rad}-P_{NBI,shinethrough}$. Previous studies on MAST [24] have shown that measurements from the Langmuir probes at the inner and outer strike points is a good proxy for the total power entering the scrape off layer. The values of $P_{Sep}$, as obtained from Langmuir probe measurements, for the four beam powers measured shortly before the L-H transition are shown in table 1. These measurements of $P_{Sep}$ are used throughout the rest of the paper to estimate the particle power through the separatrix for discharges with the various injected neutral beam power. Since the discharge with 0.6MW of injected beam power was close to the power threshold, the $P_{th}$ for these 400kA discharges is estimated to be ~0.48MW.

Previous studies of power threshold on MAST [19] were performed for a 600kA LSND discharge with slightly different shaping to the 400kA discharges discussed here. These studies found marginal L-H dithers for an ohmic discharge and a clear transition to H-mode on injection of 0.3MW of beam power corresponding to a $P_{th}=0.8+/-0.1$MW. These 600kA discharges have ohmic powers of ~0.35MW as compared to ~0.175MW for the 400kA discharges. In addition these discharges had an on axis $B_T$ of ~0.60T compared to 0.48T for the 400kA discharges and since $P_{th} \sim B_T^{0.8}$ the relative power thresholds determined for the two sets of discharges are quite reasonable.

## 4. Impact of n=6 RMPs on L-H transition for a 400kA discharge

The impact of varying RMP coil current on a typical discharge is shown in figure 3. This discharge has 1.5MW of injected neutral beam power and a $P_{Sep} \sim 1.04MW$ and hence is at least a factor of 2 above the L-H power threshold. In this case RMPs are applied in an n=6 configuration, with coil currents of 0.6kA, 1.0kA and 1.4kA. The line integral density is held constant from 300ms and the L-H transition for all discharges occurs at a value of $n_e dl = 1.25 \times 10^{21} m^{-2}$. Similar density profiles from the Thomson scattering diagnostic are observed for the no RMP, $I_{RMP}=1.0kA$ and $I_{RMP}=1.4kA$ discharges up to the timing of the L-H transition. No Thomson scattering data is available for the $I_{RMP}=0.6kA$ discharge, however a similar time evolution of the line integral density is observed from the interferometer up to the timing of the L-H transition. A particle 'pump out' caused by the RMPs is evidenced by an increased gas refuelling rate with increasing RMP coil current as shown in figure 3(b). Once in H-mode, the gas fuelling is turned off by the feedback system. A strong increase in ELM frequency (~doubling) is observed in the discharges with applied RMP. The impact on the timing of the L-H transition is very large with delays of ~13,120 and 130ms for 0.6kA, 1.0kA and 1.4kA respectively corresponding to a range ~0.4-4 energy confinement times.

No significant difference is observed in the $n_e$ or $T_e$ profiles which could account for this delay in transition, as shown in figure 4 (a) and (d). However, there are changes to the carbon emissivity, toroidal velocity and $T_i$ as shown in figure 4 (b), (c) and (e) respectively. This data is obtained from measurements of charge exchange between injected beam ions and carbon impurities [25]. Measurements of the charge exchange emission at the separatrix are not well resolved due to the instrument function and strong background emission, however, measurements a few cm inside the last closed flux surface are reliable.

An estimate of the radial electric field in the edge region of the charge exchange measurements is shown in figure 4 (g). This radial electric field is given by $E_r = dP_i/dr - v_\theta B_\phi + v_\phi B_\theta$ and contributions from the ion pressure gradient and toroidal velocity terms are used for the results shown. The poloidal velocity contribution to $E_r$ is not known, as poloidal velocity was not measured for these discharges, however this contribution is likely to be small as previous measurements of poloidal velocities on MAST have indicated small values [26]. As the RMP intensity is increased it may be seen that the impurity density is reduced. This is understandable as the RMPs cause a particle 'pump out' for all species and the feedback on electron density causes increased refuelling by deuterium gas puffing. A measurement of the absolute carbon density, as shown in figure 4(f), has been obtained from RGB[27] shows a decrease in peak carbon density from $3 \times 10^{17} m^{-3}$ at 1.4kA to $1.8 \times 10^{17} m^{-3}$ at 1.0, 0.6kA to $1.2 \times 10^{17} m^{-3}$ at 0.0kA. Reduced plasma velocity with increasing RMP field, as shown in figure 4(c), is typical of MAST plasmas[28]. The reduction in $T_i$ is less well understood, especially as no impact is seen on $T_e$, but could be due to the link between ion temperature and velocity. The depth of the radial electric field well, as shown for the edge in fig 4(g), decreases (becomes more positive) for increasing RMP intensity. This change of the edge radial electric field with RMP is largely due to the effects of changes in carbon temperature and density, since the toroidal velocity in this region is similar for these discharges and has a similar contribution to $E_r$.

Edge radial electric field is widely theorised to be a strong determinant of the L-H transition due to its' impact on flow shear which in turn suppresses turbulence causing a build-up of the pedestal [29]. This observation of changes in $E_r$ on MAST due to RMP leading to an impact on the L-H transition is consistent with $E_r$ measurements from ASDEX upgrade [9], reduction in toroidal velocity NSTX [15] and previous measurements of $E_r$ in 600kA discharges on MAST [16]. It should be noted that these previous measurements on MAST showed the reduction in $E_r$ was due to Lorentz ( $v \times B$ ) contribution of He$^+$ ions, while the results discussed here show the impact on $E_r$ due to the pressure gradient term for Carbon ions. A further set of results from MAST [30] were obtained from Gundestrup probe measurements. These results show evidence of a similar reduction in the depth of radial electric field well due to changes in flow measurements on application of RMP for a Connected Double Null (CDN) discharge. It should be noted that the radial electric field is the same for whichever ion species examined, however the relative contributions of the Lorentz component and pressure gradient component to that radial electric field can vary depending on the species.

### 5. Impact on L-H transition of different applied 'n'

The impact of RMPs of different 'n' number on the transition is shown in figure 5. Discharges are examined with 0.9MW, 1.2MW and 1.5MW of injected NBI power, with no applied RMP and applying n=2,3,4 and 6 perturbations, in all cases with 1.4kA of coil current. An estimate of the energy confinement time for the no applied RMP discharges of ~30ms is obtained from EFIT.

The results of these comparisons show that the n=2 has the greatest impact on H-mode access, completely suppressing the L-H transition in the 1.5MW discharge. These results indicate an increase in power threshold of at least 100% for these n=2 discharges. The n=2 RMP discharges on MAST have a much greater core plasma braking than higher 'n'. For comparison the toroidal rotation braking of these discharges at a time just before the L-H transition is shown in figure 6.

For the n=3 and 4 cases with 1.5MW of injected power there is a delay in the L-H transition time of 1-2$\tau_E$, even though at this $P_{NBI}$ the $P_{Sep}$ is at least a factor of 2 above the $P_{th}$ required for H-mode access with no applied RMP. Decreasing the beam power to 1.2MW, such that the discharges are at least a factor of 1.6 above the no RMP $P_{th}$, the timing of the L-H transition is delayed by $\sim3\tau_E$. No sustained H-mode access is obtained for shots in n=3 and n=4 at injected NBI powers of 0.9MW. This implies an increase in power threshold of at least 20% above the no RMP $P_{th}$.

Application of RMPs with the same RMP coil current in an n=6 configuration causes a larger impact on the L-H transition than n=3,4. For the 1.5MW discharge the timing of the L-H transition is delayed by $\sim3\tau_E$, although once the transition occurs the H-mode is stable. For the 1.2MW discharge a short H-mode period is obtained followed closely by a back transition, indicating that for this level of applied RMP the plasma is close to its L-H power threshold. Hence the power threshold is increased by at least 60% at this RMP field for n=6 discharges.

Expressing the increased power requirements to access H-mode in terms of injected power, an increase of 50% in external heating is required for n=3 and 4 discharges and an increase of 100% in external heating is required for n=6 discharges. Comparing in terms of power through the separatrix, the respective values of 20% and 60% are considerably lower. This is due to the relative magnitudes of radiated power, ohmic heating power and neutral beam shine through fraction.

The figure of merit for ELM mitigation is the achieved increase in ELM frequency due to the RMP versus the cost, which is the reduction in energy confinement due to the RMP induced particle pump out. The application of 1.4kA of RMP coil current for n=3,4 and 6 results in an increase from $f_{ELM}\sim25Hz$ to $f_{mitigated}\sim65Hz$ at 1.5MW, and from $f_{ELM}\sim15Hz$ to $f_{mitigated}\sim55Hz$ in the 1.2MW case. Both sets of mitigated discharges show a drop in energy confinement of $\sim$20-30%.

The results from pulses already shown in figure 5, as well as from further discharges with RMP currents of 0.6kA and 1.0kA are shown in figure 7. Each point in this figure represents either a discharge that either remained in L-mode, symbolised by a diamond, or transitioned to H-mode, symbolised by a circle. The delay to the timing of the L-H transition is not taken into account, so even discharges with delays as long as $7\tau_E$ are considered as successful transitions. The top row shows discharges as a function of injected neutral beam power and current in the RMP coils. The bottom row shows the same discharges in terms of $P_{Sep}$ relative to the no applied RMP power threshold and $b^r_{res}$. For definition of $b^r_{res}$ see page 47 of [31].

The data presented in figure 5 compared different applied RMP 'n' numbers at the same $I_{RMP}$, however, the magnitude of $b^r_{res}$ varies with 'n' number and is a better metric for comparison as it is machine independent. ERGOS [32] vacuum simulations were run showing the magnitude of $b^r_{res} =[0.78,0.66,0.81,0.85] \times 10^{-3}$ at $=0.95$ for n=[2,3,4,6]. There is a large n=2 intrinsic field component, taking this into account $b^r_{res}=0.94\times10^{-3}$ at $=0.95$ for n=2. The relative impact of n=3,4 and 6 may be compared using similar regions of $b^r_{res}$ shown as the grey boxes of figure 7(d-f). This shows that n=6 RMP has a greater impact on the L-H power threshold and also indicates that n=3 may have a slightly higher impact than n=4.

For extrapolation to future devices it is important to determine how the power threshold varies with $b^r_{res}$. A study presented in [17] has shown that the mitigated ELM frequency increases linearly above some minimum $b^r_{res}$ value. In particular that study showed that for the n=4 case the ELM frequency is the same as an unmitigated discharge at $b^r_{res}\sim0.5x10^{-3}$ and increases rapidly such that $f_{mitigated}\sim3\times f_{natural}$ at $b^r_{res}\sim1.0x10^{-3}$ . The power thresholds for n=4 RMP at both $I_{ELM}=1.0kA$ and $I_{ELM}=1.4kA$ may be estimated from figure 7(b) and (e). The L-H transition is suppressed at $b^r_{res} \sim 0.85 \times 10^{-3}$ when $P_{Sep} = 1.2 * P_{Thr}$, however, the L-H transition occurs at the same power level when the perturbed field is reduced to $b^r_{res} \sim 0.6 \times 10^{-3}$. This indicates that there is a minimum threshold in $b^r_{res}$ below which there is little increase in power threshold and above which the power threshold increases quite rapidly. The implication of this is that if a large ELM mitigation is required, such as a factor of 10, to satisfy material lifetime constraints for plasma facing components a very large increase in L-H power threshold could result.

## 6. Conclusions

Application of RMP to MAST discharges close to the L-H power threshold delays the L-H transition at low applied field and prevents of the transition at high field. These effects occur for a wide range of discharges including 600kA sawtoothing discharges at constant gas fuelling and 400kA sawtooth free discharges when gas fuelling is used in feedback to keep line integral density constant. Application of RMP during density feedback does not affect $n_e$ or $T_e$ profiles however there is a significant braking of the plasma. For an n=6 case examined in detail, changes in the carbon ion temperature profile are shown to impact on the radial electric field.

In order to quantify the increased power threshold, discharges where the L-H transition is prevented due to applied RMP are compared with the determined power threshold for no applied RMP discharges. The particular case of a 400kA  lower single null discharge with a $b^r_{res}$ sufficient to cause a mitigated frequency of $f_{ELM}/f_{Natural} \sim 3$ was used as a test case. Varying the toroidal perturbation of the applied field the required neutral beam power for H-mode access was 0.9MW for n=3 and 4, 1.2MW for n=6 and >1.5MW for n=2 compared with <0.6MW for no applied RMP. Expressed in power threshold using estimates of $P_{Sep}$ from Langmuir probe data, this corresponds to increases of at least 20% for n=3 and 4, 60% for n=6 and 100% for n=2. The increase in power threshold increases with applied field above a certain minimum value, similar to the trend observed for increase of ELM frequency with RMP. The combination of these scalings implies that if large mitigation factors are required on future devices to satisfy either the material lifetime constrains of plasma facing components or to prevent impurity accumulation then large increases in external heating power will be required to ensure H-mode access.

**Acknowledgements**

This work was funded by the RCUK Energy Programme under the grant EP/I501045. To obtain further information on the data and models underlying this paper please contact PublicationsManager@ccfe.ac.uk. The views expressed herein do not necessarily reflect those of the European Commission.

**References**

**Tables:**

| $P_{NBI}$ (MW) | $P_{Sep}$ (MW) |
|---|---|
| 1.5 | 1.04 |
| 1.2 | 0.77 |
| 0.9 | 0.57 |
| 0.6 | 0.48 |

**Table 1 – Injected neutral beam power and corresponding power through the separatrix as measured by Langmuir probes for 400kA LSND discharges. This value of $P_{Sep}$ is used for calculation of power threshold in equivalent discharges with applied RMP.**

**Figures:**

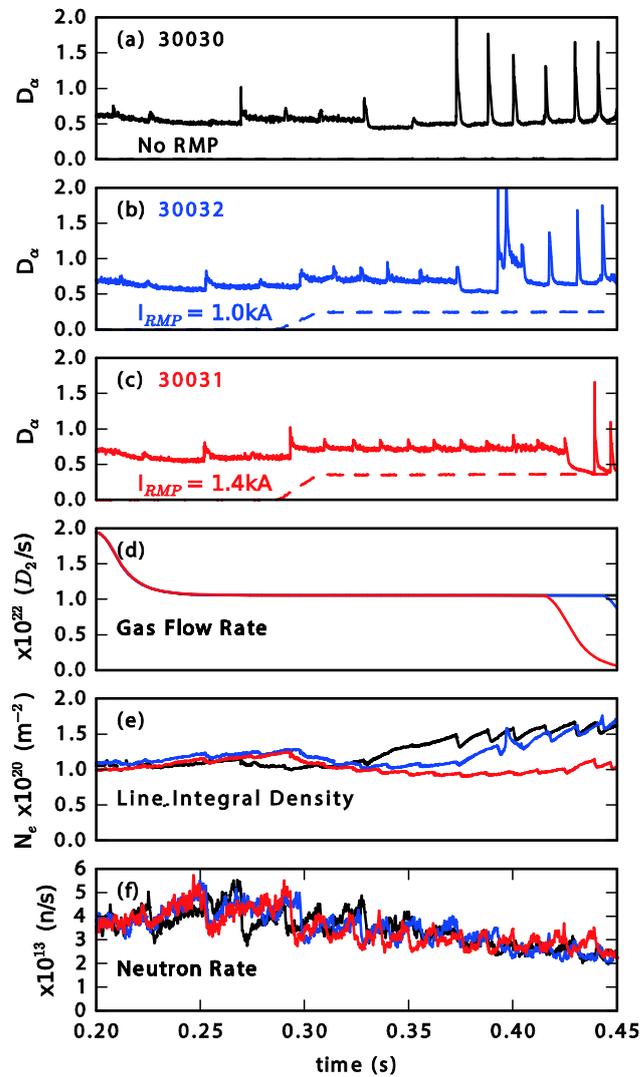

**Figure 1 – Impact of RMPs on L-H transition for a 600kA discharge with constant gas fuelling.**
**(a-c) Midplane Dᵧ emission (d) Deuterium gas fuelling, all fuelling in this discharge is low field side (e) line integral density from interferometer measurements (f) neutron rate.**

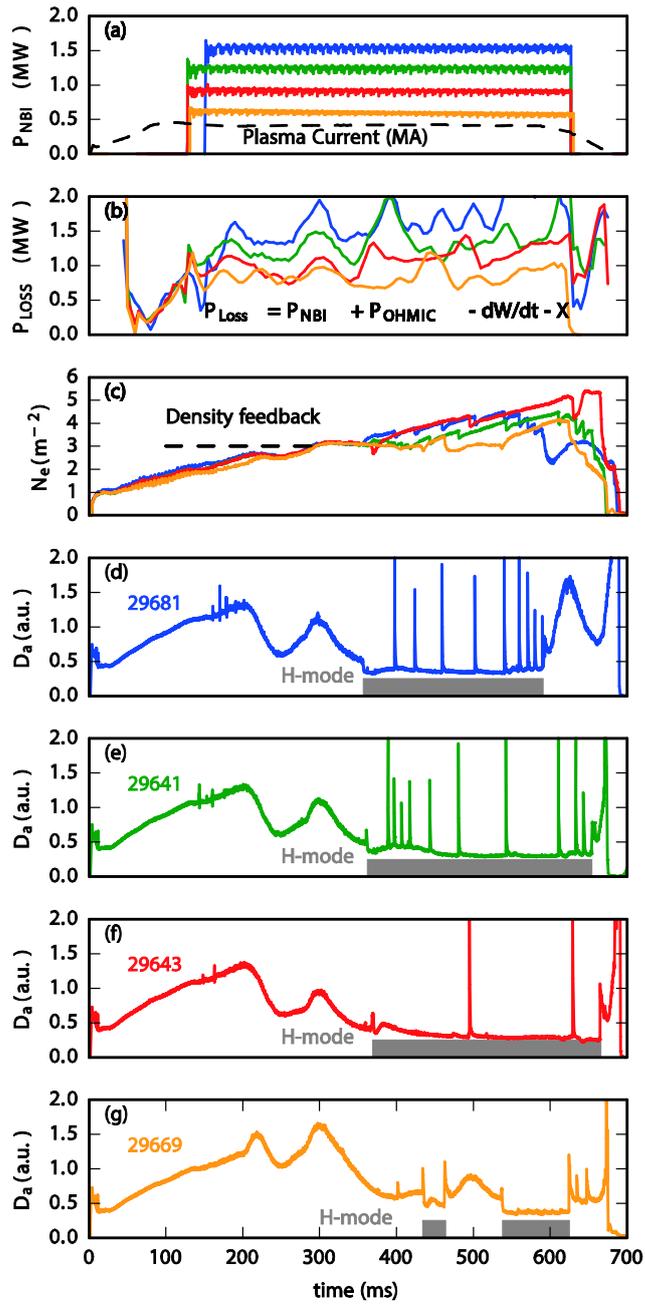

**Figure 2 – A beam power scan performed for 400kA lower single null discharges to determine the threshold power for L-H transition. The plots show (a) injected neutral beam power, (b) loss power estimated from equilibrium reconstruction excluding neutral beam shine through and radiative losses, (c) line integral density showing the density feedback level and (d-g) D⌐ emissions for each pulse.**

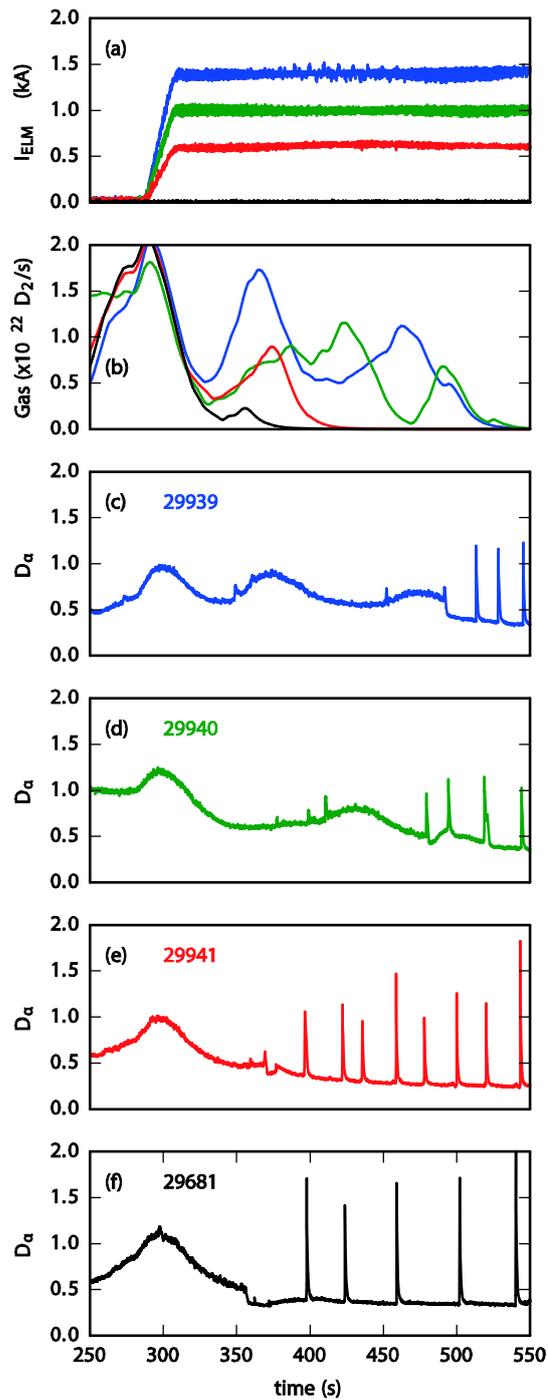

**Figure 3 – Impact of increasing RMP coil intensity, in n=6 configuration, on timing of the L-H transition for a 400kA discharge with 1.5MW of injected neutral beam power. (a) RMP coil current (b) D$_2$ gas refuelling rate in density feedback (c-f) D-alpha emission as measured at the midplane.**

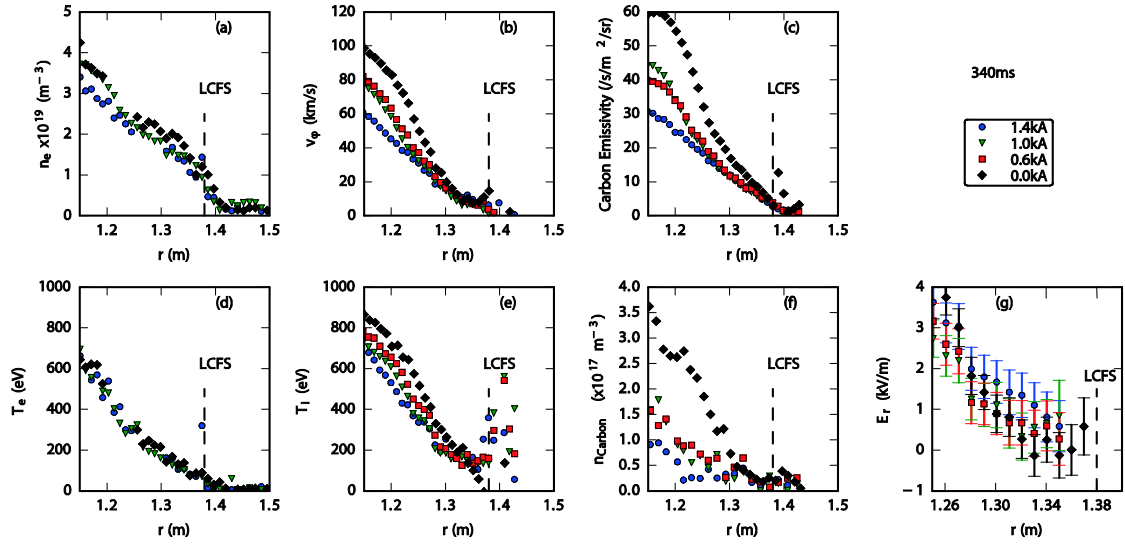

**Figure 4 – Impact of RMPs on the plasma profiles for the pulses shown in figure 3 at 340ms before the L-H transition. There is no Thomson scattering data available for the 0.6kA pulse. (a) electron density (b) carbon emissivity (c) toroidal velocity (d) electron temperature (e) carbon ion temperature (f) Edge radial electric field (has a different radial axis with respect to the other profiles).**

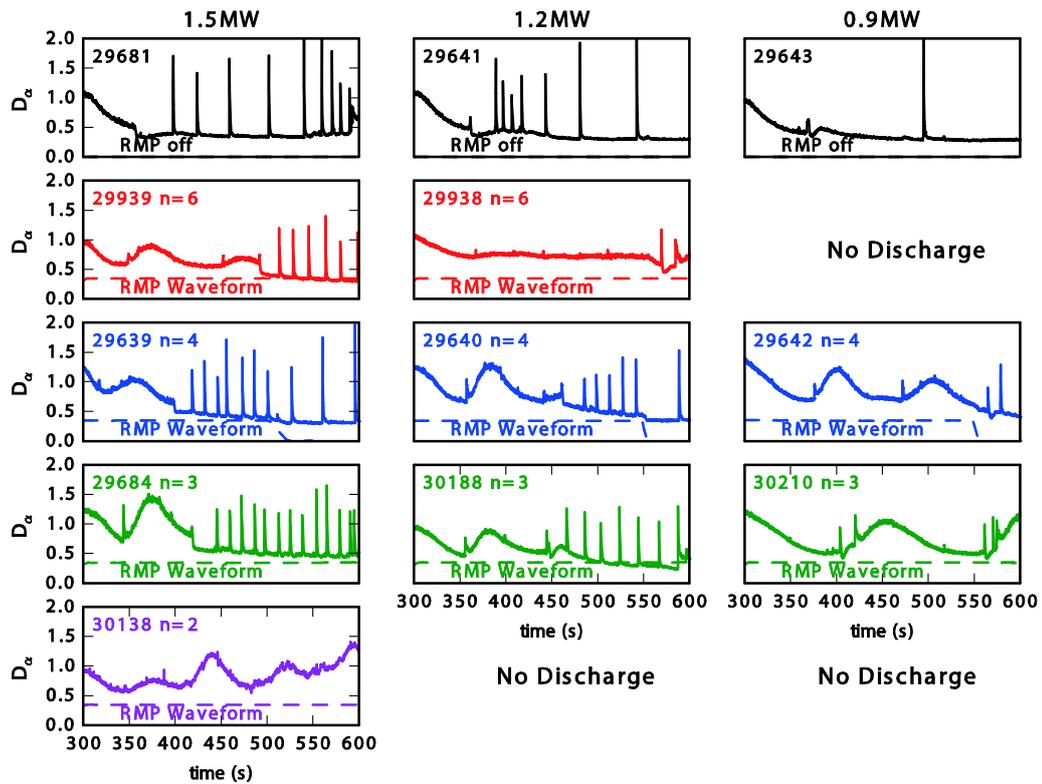

**Figure 5** – The measured Dᴦ emission for a range of discharges. The columns correspond to injected neutral beam powers of $P_{NBI}$=1.5, 1.2 and 0.9MW respectively. The rows correspond to different RMP configurations, no applied RMP in the top row and applied n=6,4,3,2 perturbations in the subsequent rows. The RMP coil current waveforms are also shown, all discharges with RMP are at an intensity of $I_{RMP}$=1.4kA.

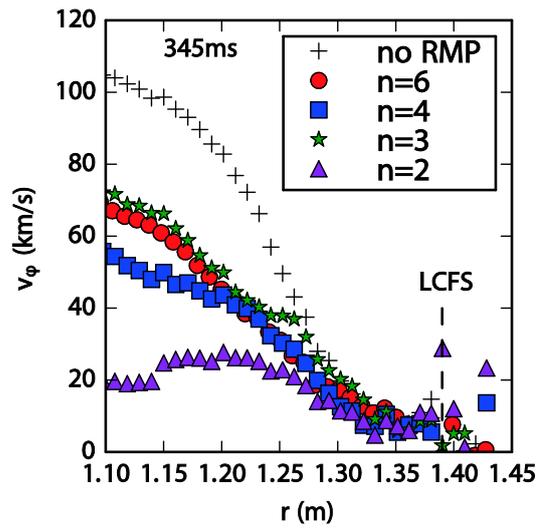

**Figure 6 – Radial profiles of toroidal velocity for 1.5MW discharges shown in first column fig bb. Measurements are taken at 345ms, before the timing of the L-H transition in the no applied RMP discharge.**

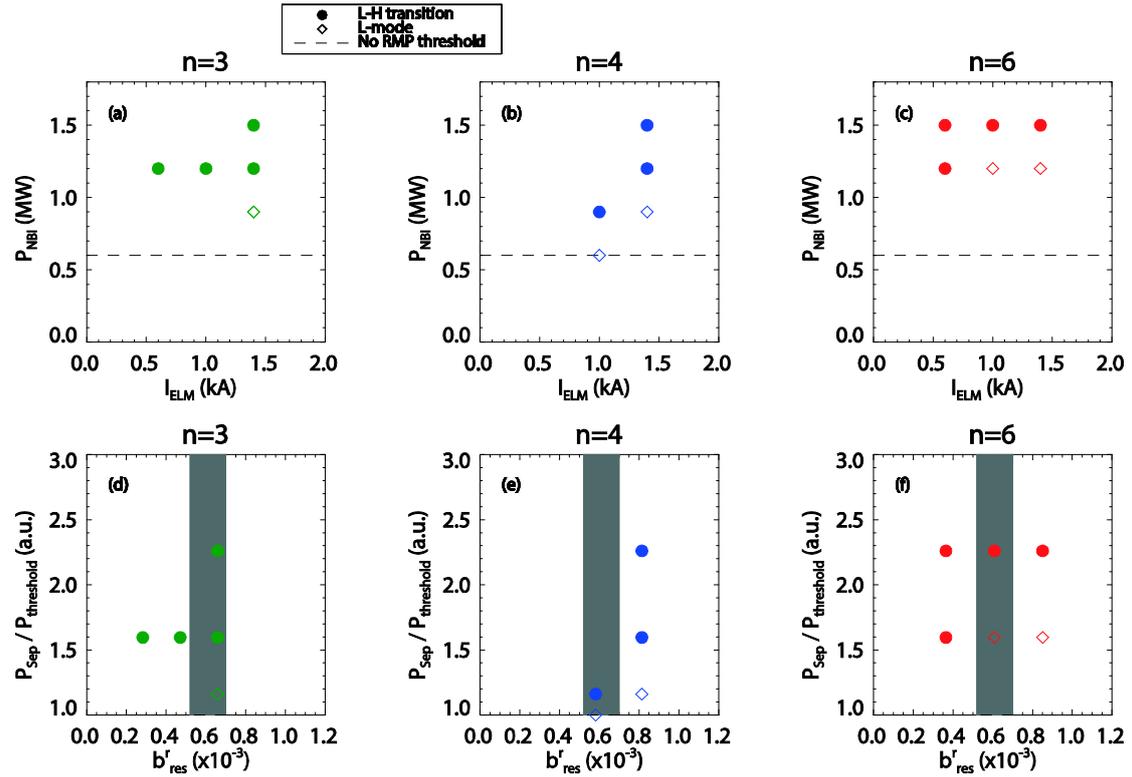

**Figure 7** - Map of discharges showing impact of RMP on H-mode accessibility. Circles indicate discharges that did transition to H-mode, irrespective of the delay in L-H transition time. Diamonds show discharges that remained in L-mode. (a-c) Injected neutral beam power versus RMP coil current for n=3,4,6 (d-f) Estimated $P_{Sep}$ normalised to $P_{threshold}$ versus $b^r_{res}$. A region of constant $b^r_{res}$ is highlighted across the three 'n' numbers for comparison.